\documentclass[10pt,letterpaper,twocolumn]{article} 

\usepackage{times,mathptmx}
\usepackage{ol2}
\usepackage{amsmath,amssymb}

\begin{document}

\twocolumn[ 

\title{Coherence properties of Kerr frequency combs}


\vskip-4mm

\author{Miro Erkintalo$^{*}$ and St\'ephane Coen}

\address{
Physics Department, The University of Auckland, Private Bag 92019, Auckland 1142, New Zealand\\
$^*$Corresponding author: m.erkintalo@auckland.ac.nz
}

\begin{abstract}%
  We use numerical simulations based on an extended Lugiato-Lefever equation (LLE) to investigate the stability
  properties of Kerr frequency combs generated in microresonators. In particular, we show that an ensemble average
  calculated over sequences of output fields separated by a fixed number of resonator roundtrips
  allows the coherence of Kerr combs to be quantified in terms of the complex-degree of first-order coherence. We
  identify different regimes of comb coherence, linked to the solutions of the LLE. Our approach provides a
  practical and unambiguous way of assessing the stability of Kerr combs that is directly connected to an
  accessible experimental quantity.
\end{abstract}


 ] 

\noindent Kerr microresonators offer a promising route towards miniaturization of the frequency comb
technology, which has led to intense investigations of ``Kerr combs'' over the last few years
\cite{kippenberg_microresonator-based_2011, delhaye_optical_2007, savchenkov_tunable_2008, delhaye_octave_2011,
okawachi_octave-spanning_2011, ferdous_spectral_2011, papp_spectral_2011, herr_universal_2012,
grudinin_frequency_2012, saha_modelocking_2013}. Despite being excited by continuous-wave (cw) light, these combs
are still in some conditions underlined by ultrafast pulses in the form of localized dissipative structures known as
temporal cavity solitons (CSs) \cite{wabnitz_suppression_1993, leo_temporal_2010, jang_ultraweak_2013}. This was
initially suggested in \cite{leo_temporal_2010} and subsequently demonstrated both theoretically
\cite{matsko_mode-locked_2011, coen_modeling_2013, coen_universal_2013, chembo_spatiotemporal_2013,
lamont_route_2013} and experimentally \cite{herr_mode-locking_2012}. The microresonator technology therefore also
holds potential for on-chip generation of high repetition-rate pulse trains.

Stability is a key factor for many applications of frequency combs. Quantifying the noise properties of
microresonator-based combs has, however, proven difficult. Radio-frequency measurements are hindered
by large resonator free-spectral ranges (FSRs) \cite{papp_spectral_2011, herr_universal_2012}. Additional optical
heterodyning relaxes this limitation but is rather complex to implement \cite{herr_universal_2012,
herr_mode-locking_2012}. These two methods do not however easily reveal how noise varies across the comb and offer
limited insights on time-domain properties. Comb coherence has also been inferred from autocorrelation-type temporal
signals \cite{ferdous_spectral_2011, papp_spectral_2011}, yet these are average measurements that are prone to
misinterpretations \cite{rhodes_pulse-shape_2013}. An alternative method has been used in the
context of supercontinuum (SC) frequency combs. Two independently generated SC were superimposed on a spectrometer,
with the resulting spectral interference pattern providing direct evidence of the SC coherence
\cite{bellini_phase-locked_2000}. Of course, the visibility of the interference fringes is simply the modulus of the
wavelength-dependent complex degree of first-order coherence, which has been exploited to elucidate the mechanisms
of SC coherence degradation both numerically \cite{dudley_coherence_2002, dudley_supercontinuum_2006} and
experimentally \cite{gu_experimental_2003, lu_generation_2004}. In this Letter, we theoretically examine Kerr comb
stability in light of this definition. Specifically, an ensemble average computed over spectral snapshots of the
simulated intracavity field separated by a fixed number of resonator roundtrips allows the wavelength-dependent
degree of coherence to be calculated. We report that the different comb dynamical regimes, previously identified in
\cite{coen_universal_2013}, exhibit qualitatively different coherence characteristics.

In our work, Kerr comb dynamics is modeled with a generalized mean-field Lugiato-Lefever equation (LLE)
\cite{coen_modeling_2013, coen_universal_2013, lamont_route_2013, chembo_spatiotemporal_2013},
\begin{align}
  \label{LL}
  t_\mathrm{R}\frac{\partial E(t,\tau)}{\partial t} =& \bigg[ -\alpha - i \delta_0
    + i L\sum_{k\geq 2}\frac{\beta_k}{k!}\left(i\frac{\partial}{\partial \tau}\right)^k \nonumber \\
  & \qquad + i\gamma L |E|^2 \bigg] E + \sqrt{\theta}E_\mathrm{in}\,.
\end{align}
Here $t$ is the slow time describing the evolution of the intracavity electric field envelope $E(t,\tau)$ over
subsequent resonator roundtrips while $\tau$ is a fast time defined in a reference frame traveling at the group
velocity of the driving field with cw amplitude $E_\mathrm{in}$. $t_\mathrm{R} = \mathrm{FSR}^{-1}$ is the roundtrip
time, and $\alpha = (\alpha_\mathrm{i}L+\theta)/2$ describes the total linear cavity losses with $\alpha_\mathrm{i}$
the internal absorption, $\theta$ the coupler power transmission coefficient, and $L$ the resonator roundtrip
length. Defining $\phi_0$ as the linear phase-shift acquired by light at the driving frequency over one roundtrip,
$\delta_0 = 2\pi l - \phi_0$ measures the phase detuning of the driving field with respect to the closest linear
resonance (with order~$l$). The $\beta_k$'s are the Taylor series expansion coefficients of the resonator
propagation constant about the angular frequency $\omega_0$ of the driving field and $\gamma$ is the nonlinearity
coefficient. Self-steepening was verified not to play a significant role with our parameters, and can be neglected.

To analyze comb stability we integrate Eq.~\eqref{LL} using the split-step Fourier method. For the sake of
discussion, we consider a silicon nitride resonator with simulation parameters approximated from
\cite{okawachi_octave-spanning_2011} and already used in \cite{coen_modeling_2013}: loaded quality factor $Q =
3\cdot 10^{5}$; $\mathrm{FSR} = 226$~GHz; $\gamma =  1\ \mathrm{W^{-1} m^{-1}}$; $\alpha = \theta = 0.009$. The
driving power is set to $P_\mathrm{in} = |E_\mathrm{in}|^2=755$~mW. We emphasize that the choice of parameters is
immaterial: the LLE can be normalized as in \cite{leo_temporal_2010, coen_universal_2013}, extending our analysis to
arbitrary configurations. Accordingly, we refer below to the normalized detuning $\Delta = \delta_0/\alpha$ and, as
the relevant dynamical timescale, to the photon lifetime $t_\mathrm{ph} = Q/\omega_0 = t_\mathrm{R}/(2\alpha)$. Here
$t_\mathrm{ph} \simeq 56\,t_\mathrm{R}$. We have numerically explored the stability characteristics of all four
classes of comb solutions of the LLE \cite{coen_universal_2013}. For increasing values of $\Delta$, these are: (i)
stable modulation instability (SMI); (ii) unstable modulation instability (UMI); (iii) unstable cavity solitons
(UCS); and (iv) stable cavity solitons (SCS). For each detuning $\Delta$, the simulation initial condition consists
of the cw steady-state solution [the upper state when bistable, $\Delta>\sqrt{3}$] superimposed with a quantum noise
seed of one photon with random phase per mode. To avoid transients, we simulate the intracavity dynamics over one
million roundtrips (more than 17,000~$t_\mathrm{ph}$) before gathering stability statistics. We
have verified this approach to yield qualitatively identical results as the more realistic (but more computationally
intensive) method of ramping $\Delta$ to bring the driving laser into resonance
\cite{herr_mode-locking_2012, lamont_route_2013}. Note that we do not put noise on the driving
field: the comb coherence we quantify here represents the intrinsic nonlinear cavity dynamics. Technical noise on
the pump will further degrade the coherence. A detailed study will be presented elsewhere, but we have found this
degradation to be relatively minor even with noise levels well above shot-noise.

\begin{figure}[t]
  \includegraphics[width = \columnwidth]{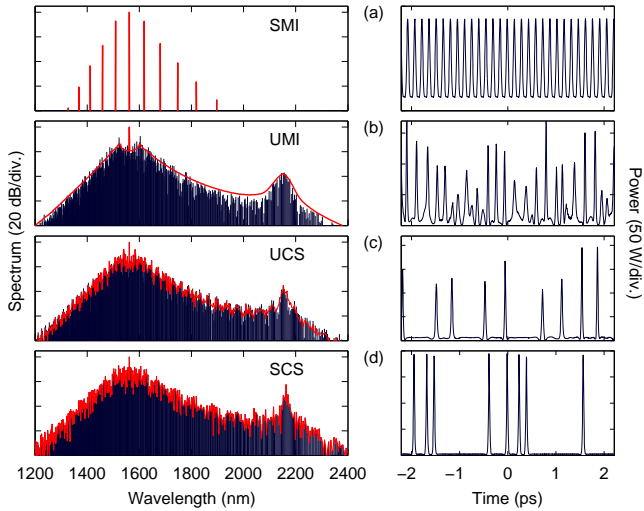}
  \caption{\small Illustrative simulation results for $\Delta = -0.5$, 3, 4, and~6 [from (a) to (d)].
    Left: single roundtrip snapshot (black) and 1-million roundtrip averaged (red) spectra. Right: single-roundtrip
    snapshot of the corresponding temporal intensity profile.}
  \label{spectra}
  \vskip -4mm
\end{figure}

Figure~\ref{spectra} shows simulated spectral and temporal characteristics illustrative of each regime of LLE
solutions. The left panels show a spectral snapshot corresponding to a single roundtrip (black) as well as the
\emph{mean} spectrum averaged over 1 million roundtrips (red); the right panels show a snapshot of the time-domain
intracavity field. Figure \ref{spectra}(a) corresponds to stable MI, obtained with $\Delta=-0.5$. The spectrum is
dominated by ``primary'' comb lines separated by multiple FSRs \cite{chembo_modal_2010}. The time-domain profile is
a periodic pattern that repeats every roundtrip: the comb is stable. Correspondingly, the mean spectrum is identical
to that of a single roundtrip. For an increased cavity detuning the MI solution becomes unstable
\cite{coen_universal_2013}. Typical simulation results are shown in Fig.~\ref{spectra}(b) for $\Delta = 3$. The mean
spectrum is smooth, and contains a broad dispersive wave (DW) pedestal at 2140~nm \cite{coen_modeling_2013}. We note
that the mean spectrum closely resembles the experimental measurement in \cite{okawachi_octave-spanning_2011}. The
comb is however unstable, and the smoothness of the spectrum arises from ensemble averaging of fluctuating,
highly-structured single-roundtrip spectra \cite{matsko_chaotic_2013}. Rapid fluctuations are also apparent in the
time-domain snapshot. Above the upswitching point, $\Delta_\uparrow \simeq 3.2$ in our case, the nature of LLE
solutions changes from MI to CSs \cite{coen_universal_2013}. Figure~\ref{spectra}(c) shows results for $\Delta = 4$
in the unstable CS regime, characterized by breathing CSs \cite{matsko_excitation_2012, leo_dynamics_2013,
coen_universal_2013} as alluded to by the time-domain snapshot. Here the breathing period is $3\,t_\mathrm{ph}$ (or
$167\,t_\mathrm{R}$) for all CSs, though in general they oscillate out-of-sync. The mean spectrum is no longer
smooth but exhibits sharp features. Finally, further increasing the detuning allows for the excitation of stable CSs
\cite{coen_universal_2013, chembo_spatiotemporal_2013, lamont_route_2013, herr_mode-locking_2012}.
Typical characteristics are shown in Fig.~\ref{spectra}(d) for $\Delta = 6$. The mean spectrum is identical with an
arbitrary selected snapshot, and both exhibit complex structure owing to the multiple CSs populating the cavity.
Naturally for a single CS the spectrum is smooth \cite{coen_modeling_2013}. We find that the number of solitons
generally varies with the input noise-seed, yet their characteristics remain the same. They all drift with the same
small constant velocity in the $\tau$-reference frame due to high-order dispersion \cite{leo_nonlinear_2013}.
Also, they do not experience \emph{relative} motion, which ensures the stability of the comb. Extra sources of noise may however affect this.

\begin{figure}[t]
  \includegraphics[width = \columnwidth]{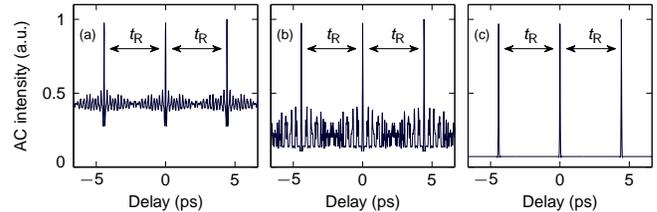}
  \caption{\small Simulated AC traces for the (a) unstable MI, (b) unstable CS, and (c) stable CS regimes.
  The cavity detuning values correspond to those in Figs.~\ref{spectra}(b)--(d).}
  \label{ACs}
  \vskip -4mm
\end{figure}

Little insights can be obtained on comb stability based on the mean spectrum. The same applies to ensemble-averaged
temporal measurements, as already mentioned in \cite{herr_mode-locking_2012}. To highlight this point, we plot in
Figs.~\ref{ACs}(a)--(b) the averaged intensity autocorrelation (AC) traces corresponding to the unstable MI and CS
regimes of Figs.~\ref{spectra}(b)--(c), respectively. The AC traces are displayed over a temporal span of three
roundtrips, and were computed from 5000 consecutive roundtrips. For comparison we also plot, in Fig.~\ref{ACs}(c),
the AC trace of a stable comb in which a single CS populates the cavity ($\Delta = 6$). In all cases the trace
consists of sharp peaks atop a background, repeating at delays separated by the cavity roundtrip time (note that the
background in the CS regime would be smaller for larger $Q$s). For the unstable regimes the peaks correspond to
``coherence artefacts'' and should not be confused with stable pulse formation \cite{rhodes_pulse-shape_2013}. They
are periodic owing to the long photon-lifetime (high-finesse) of the cavity: the field changes slowly from
roundtrip-to-roundtrip such that for delays of a few roundtrips the AC signal is almost identical to that at zero
delay. Similar artefacts appear for time-averaged frequency-resolved optical gating measurements.

\begin{figure}[t]
  \includegraphics[width = \columnwidth]{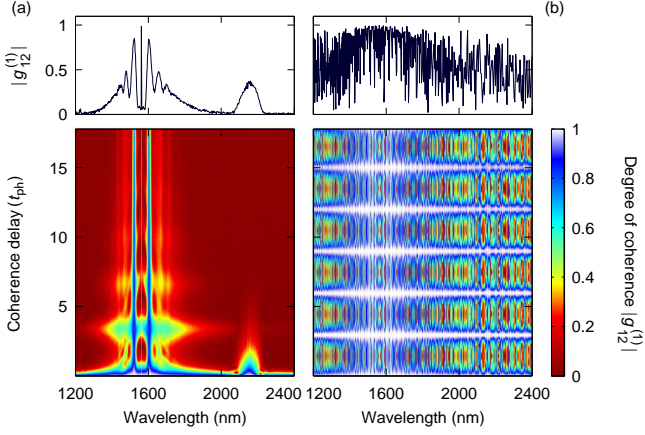}
  \caption{\small Degree of first-order coherence across simulated Kerr combs as a function of coherence delay
    (density maps) and for a two photon-lifetime delay (line plot at the top) for (a) the unstable MI and (b)
    the unstable CS regimes. The corresponding mean spectra are shown in Fig.~\ref{spectra}(b) and
    Fig.~\ref{spectra}(c), respectively.}
  \label{coherence}
\end{figure}
\begin{figure}[t]
  \includegraphics[width = \columnwidth, clip = true]{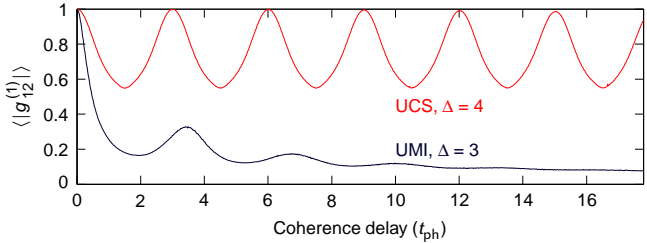}
  \caption{\small Overall degree of coherence as a function of coherence delay for the unstable MI (black) and
    unstable CS (red) regimes.}
  \label{coherence2}
  \vskip -4mm
\end{figure}
To examine the coherence of the simulated Kerr combs, we calculate the modulus of the complex degree of first-order
coherence at each wavelength $\lambda$,
\begin{equation}
  \left|g_{12}^{(1)}(\lambda, t_1-t_2)\right| =
    \frac{\left|\langle \tilde{E}^{\ast}(t_1,\lambda)\tilde{E}(t_2,\lambda)\rangle\right|}{S(\lambda)}\,.
  \label{g12}
\end{equation}
Here, $\tilde{E}$ is the complex Fourier transform of the intracavity (or output) electric field (both give the same
result), the angle brackets denote an ensemble average, and $S(\lambda) = \langle |\tilde{E}(\lambda)|^2 \rangle$ is
the mean spectrum. We evaluate $|g_{12}^{(1)}|$ at coherence delays $|t_1-t_2| = nt_\mathrm{R}$, where $n$ is a
positive integer (intermediate delays only affect the periodicity of the fringe pattern, not the visibility). In
other words, we examine correlations between intracavity fields separated by a set number of cavity roundtrips. Note
that this differs from SC coherence calculations in which all pairs of outputs are considered as the noise varies
randomly from shot to shot \cite{dudley_coherence_2002}, while for Kerr combs the sequence is deterministic and set
by the cavity dynamics. The stable MI and CS regimes have almost perfect coherence, i.e.,
$|g_{12}^{(1)}(\lambda,nt_\mathrm{R})| = 1$ for all $n, \lambda$. To the contrary, the unstable regimes exhibit a
complex dependence on both $\lambda$ and $n$ as illustrated in Fig.~\ref{coherence} for the same detunings as in
Figs.~\ref{spectra}(b)--(c). The density maps show $|g_{12}^{(1)}(\lambda,nt_\mathrm{R})|$ as a function of
wavelength and coherence delay, while the line plots at the top show the degree of coherence for a selected delay
corresponding to two photon lifetimes ($112\,t_\mathrm{R}$). As expected based on the high cavity finesse, both
regimes display strong correlations over small delays ($nt_\mathrm{R} \ll t_\mathrm{ph}$), in agreement with the
presence of coherence artifacts in the AC traces of Figs.~\ref{ACs}(a)--(b). The coherence of the unstable MI comb
[Fig.~\ref{coherence}(a)] rapidly degrades for larger delays. At $|t_1-t_2| = 112\,t_\mathrm{R} \sim
2\,t_\mathrm{ph}$ the overall coherence is almost vanished with $|g_{12}^{(1)}|\sim 0$ over most of the comb.
Residual coherence manifests only at the DW edge and for the first few primary comb lines, i.e., those experiencing
maximum parametric gain. For even larger delays the DW also becomes decorrelated, but the fundamental parametric
sidebands remain quasi-coherent. The lack of fluctuations in the vicinity of these frequencies is also corroborated
by the two sharp features in the mean spectrum plotted in Fig.~\ref{spectra}(b).

An entirely different behavior is observed in the unstable (breathing) CS regime [Fig.~\ref{coherence}(b)]. The
degree of coherence degrades much more slowly than in the UMI regime and is periodically restored for increasing
delays, here with a period of $3\,t_\mathrm{ph}$. This period is found to match with the breathing period of the
associated oscillatory CSs described above. Because the entire intracavity field repeats over the breathing period,
strong correlations, $|g_{12}^{(1)}|\sim 1$, are expected at delays multiple of that period. The
different coherence properties of the two regimes are further highlighted in Fig.~\ref{coherence2} by considering
the \emph{overall coherence} $\langle |g_{12}^{(1)}(nt_\mathrm{R})|\rangle$ averaged over the exploitable comb.
Specifically, we average $|g^{(1)}_{12}(\lambda,nt_\mathrm{R})|$ over wavelengths at which the mean comb power
exceeds $-80$~dB (relative to the driving mode, see Fig.~\ref{spectra}). The coherence oscillations of the unstable
CS regime are clearly visible on the plot (red) while the overall coherence drops rapidly for the unstable MI regime
(black), with a subsequent slow decline towards a quasi-incoherent value of $\langle
|g_{12}^{(1)}(nt_\mathrm{R})|\rangle \sim 0.1$. Note that even this latter regime exhibits decaying coherence
oscillations (with a period $3.5\,t_\mathrm{ph}$) which are reminiscent of a breathing (Hopf) instability.

\begin{figure}[t]
  \includegraphics[width = \columnwidth]{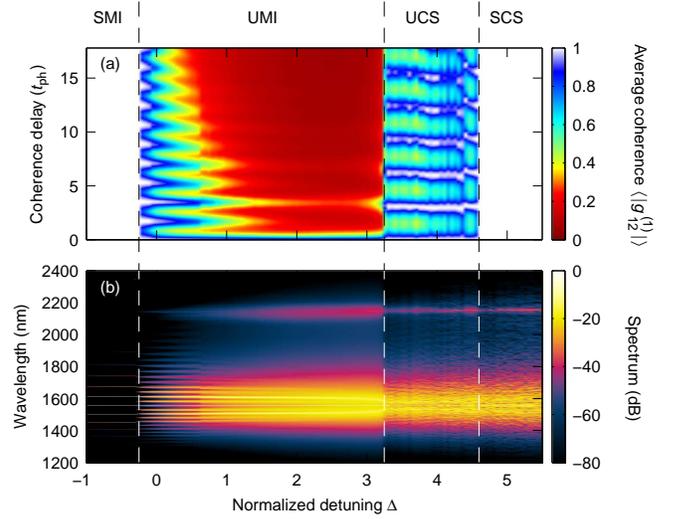}
  \caption{\small (a) Overall degree of coherence over a wide range of cavity detunings. The corresponding mean spectra are shown in (b).}
  \label{deltascan}
  \vskip -4mm
\end{figure}
In order to show that the coherence properties described above are not specific to our examples but rather are
characteristics of each representative comb regimes, we have repeated our analysis over a wide range of cavity
detunings. Our results are summarized in Fig.~\ref{deltascan}. Specifically, in Fig.~\ref{deltascan}(a) we show the
overall degree of coherence as a function of coherence delay and cavity detuning, while Fig.~\ref{deltascan}(b)
displays the corresponding mean spectrum. The four distinct classes of LLE solutions can be clearly evidenced from
both graphs, as highlighted. The coherence in the UMI regime deteriorates progressively for increased cavity
detuning, before being abruptly ameliorated on transition to the UCS regime, at a detuning corresponding to the cw
upswitching point ($\Delta_\uparrow \simeq 3.2$). In contrast, the level of coherence changes smoothly between
stable and unstable solutions of the same family (MI or CSs). Islands of varying coherence visible in the UCS regime
are due to the stochastic nature of our simulations: different configurations of CSs are excited depending on the
initial noise-seed as well as the detuning $\Delta$. In particular, a range of multiple-CS bound states emerge in
varying numbers. Local variations in the overall comb coherence follow because bound CSs appear to exhibit degraded
coherence compared to isolated CSs. This observation may be explained by differences in the Hopf threshold of bound
and isolated CS states, with the former typically occurring at a higher detuning and defining the upper boundary of
the UCS regime.

Our results have a number of important implications. On the one hand, it is clear that Kerr combs in the unstable MI
or CS regimes are undesirable owing to the spectral and temporal fluctuations intrinsic to both. On the other hand,
our results show that the unstable CS regime exhibits coherence which is greatly improved from that of MI. This
observation raises a critical warning, because past studies have associated a drop in the low-frequency radio
frequency spectrum as evidence of stable pulse formation. In light of our results, such observations could
correspond to the transition from unstable MI to \emph{unstable} CSs. The difficulty of unambiguously evidencing a
stable soliton state is further compounded by potential pitfalls in ensemble-averaged time-domain measurements.
The use of the complex degree of first-order coherence, as proposed here, alleviates all these
deficiencies, and offers a rapid route towards broadband comb characterization. We emphasize that
$|g_{12}^{(1)}(\lambda)|$ is readily accessible experimentally from the visibility of spectral interference fringes,
requiring only a standard optical spectrum analyzer and a delayed interferometer \cite{gu_experimental_2003,
lu_generation_2004}. From our results, a temporal delay of the order of a few photon lifetimes is sufficient to
discern between the different regimes. $Q$ is the dominant factor in $t_\mathrm{ph}=Q/\omega_0$. The largest delays
are thus required for crystalline resonators, which exhibit the highest $Q$-factors. Taking $Q=10^9$ yields
$t_\mathrm{ph} = 0.82~\mu\mathrm{s}$ (at 1550 nm). Such delays can be routinely obtained using a fiber-based
interferometer \cite{runge_coherence_2013}. Lower-Q resonators, like the silicon nitride device considered in our
analysis above, require much shorter delays of a few tenth of nanoseconds, readily accessible in bulk. We therefore
expect experimental measurements of Kerr comb coherence soon to follow. Finally, we note that additional insights on
the mutual coherence of distinct comb frequencies may also be obtained via the use of two-dimensional correlation
functions of second-order coherence theory \cite{genty_complete_2011, erkintalo_coherent-mode_2012}. This will be
explored in a future work.

We acknowledge support from the Marsden Fund of the Royal Society of New Zealand.

\end{document}